\documentclass[a4paper,fleqn,usenatbib]{mnras}
\usepackage{mathptmx}%uncommented 5/10
\usepackage[T1]{fontenc}
\usepackage{ae,aecompl}
\usepackage{multirow,color}
\usepackage{dcolumn}% Align table columns on decimal point
\usepackage{bm}% bold math
\usepackage{amssymb,amsmath,amsfonts,graphicx,epsf}
\usepackage{color}
\usepackage{enumerate}
\newcommand{\red}[1]{#1}

\newcommand{\bvec}[1]{\mathbf{#1}}

\title{Stability analysis of core-strahl electron distributions in the solar wind}

\author[K. Horaites et al.]{
Konstantinos Horaites,$^{1}$\thanks{E-mail: horaites@wisc.edu}
Patrick Astfalk,$^{2}$
Stanislav Boldyrev,$^{1,3}$
\newauthor
and Frank Jenko$^{2,4,5}$
\\
$^{1}$Department of Physics, University of Wisconsin -- Madison, 1150 University Avenue, Madison, WI 53706, USA\\
$^{2}$Max Planck Institute for Plasma Physics, 85748 Garching, Germany\\
$^{3}$Space Science Institute, Boulder, CO 80301, USA\\
$^{4}$Technical University of Munich, 85748 Garching, Germany\\
$^{5}$The University of Texas at Austin, Austin, TX 78712, USA\\
}

\date{Accepted XXX. Received YYY; in original form ZZZ}

\pubyear{2018}

\begin{document}
\label{firstpage}
\pagerange{\pageref{firstpage}--\pageref{lastpage}}
\maketitle

\begin{abstract}
In this work, we analyze the kinetic stability of a solar wind electron distribution composed of core and strahl subpopulations. The core is modeled by a drifting Maxwellian distribution, while the strahl is modeled by an analytic function recently derived in~\cite[][]{horaites18} from the collisional kinetic equation.    %that has recently been shown to agree with solar wind observations. 
We perform a numerical linear stability analysis using the LEOPARD solver \citep{astfalk17}, which allows for arbitrary gyrotropic distribution functions in a magnetized plasma. \red{In contrast with previous reports, we do not find evidence for a whistler instability directly associated with the electron strahl. This may be related to the more realistic shape of the electron strahl distribution function adopted in our work, as compared to previous studies.}  We however find that for typical solar wind conditions, the core-strahl distribution is unstable to the kinetic Alfv\'en and magnetosonic modes. The maximum growth rates for these instabilities occur at wavenumbers $k d_i \lesssim 1$ (where $d_i$ is the ion inertial length), at moderately oblique angles of propagation, thus providing a potential source of kinetic-scale turbulence. We therefore suggest that if the whistler modes are invoked to explain anomalous scattering of strahl particles, these modes may appear as a result of nonlinear mode coupling and turbulent cascade originating at scales $kd_i \lesssim 1$.
\end{abstract}

% Select between one and six entries from the list of approved keywords.
% Don't make up new ones.

\maketitle

\section{Introduction}\label{intro_sec}

The low Coulomb collisionality of the solar wind allows electron velocity distribution functions (eVDFs) to persist in a non-thermal state. Solar wind eVDFs exhibit suprathermal populations \citep[e.g., ][]{feldman75}, in addition to the Maxwellian core, which carry free energy that may drive kinetic instabilities. At 1 AU, suprathermal electrons with energies between $\sim$10 eV and $\sim$1 keV are typically divided into two components: the halo and strahl \citep[e.g.,][]{pilipp87}. The halo component is relatively isotropic, and is often modeled in velocity space as a kappa distribution \citep[e.g.,][]{maksimovic05}. The strahl, on the other hand, forms a narrow beam in velocity space that is aligned with the local magnetic field $\bvec{B}$, and flows anti-sunward. The strahl component is believed to be composed of ``thermal runaway'' particles \citep{gurevichistomin79}, which are seen in collisionless plasmas in the presence of a temperature gradient, that are focused into a beam as they move through a spatially weakening magnetic field \citep[e.g.,][]{scudderolbert79}.

The field-parallel skewness of solar wind eVDFs, and the heat flux associated with it, has spurred the study of so-called ``heat flux instabilities'' \citep[e.g.,][]{forslund70, gary75}. In these models, the field-parallel bulk velocity of the core population drifts slightly sunward relative to the ions. This drift offsets the anti-sunward flow of the heat flux-carrying particles so that the net parallel current is nearly zero. 
%The zero-current constraint is consistent with observations \citep{feldman75}. 
If the model parameters are adjusted so that the heat flux and associated core drift amplitude both increase, eventually an instability threshold may be reached for a particular wave mode. 
\red{\cite{gary75} modeled the eVDF as a combination of two drifting Maxwellians, which respectively represented the core and halo populations. They found these core-halo distributions to be unstable to Alfv\'en, magnetosonic, and whistler mode fluctuations, with instability regimes depending on the wavenumber.}
 Empirically, it has been shown that the electron heat flux $q$ may be constrained by thresholds imposed by the whistler as well as the kinetic Alfv\'en \citep{gary99,wilson13,tong18} heat flux instabilities; \red{these thresholds are derived assuming a core-halo model eVDF}. We note that the core drift and heat flux appear to also be mediated by the Coulomb collisionality of the solar wind, as was shown in large statistical studies of {\it Wind} satellite data \citep{bale13,pulupa14}.

Kinetic instabilities have garnered particular interest in studies of the strahl population, as they may provide a source of  \red{scattering} of the strahl beam. 
%Kinetic instabilities have garnered particular interest in studies of the strahl population, as they may provide a source of  ``anomalous diffusion'' of the strahl beam. 
%Whistler waves can resonate with strahl electrons that meet the cyclotron resonance condition, and pitch-angle scatter these particles. 
Such a source was called for by \cite{lemonsfeldman83}, who incorporated Coulomb collisions into a collisionless model of strahl formation. After comparing with data from the {\it IMP 8} satellite, it was claimed that pitch-angle scattering by Coulomb collisions alone could not account for the observed angular width of the strahl. \red{Following e.g., \cite{pilipp87}, we refer to the non-Coulombic scattering that is sometimes invoked to explain the strahl width as ``anomalous diffusion'' or ``anomalous scattering''.} Recently, \cite{graham18} inferred the presence of a strahl-scattering process that occurs over distance, by showing that for 1 AU data measured near SEP events, the angular width of the strahl population is correlated with the length of the interplanetary magnetic field lines that stretch back to the coronal base. \red{Beyond anomalous diffusion, wave-particle scattering of the strahl has been suggested by some authors to be a potential source of the nearly-isotropic halo distribution. For instance, \cite{maksimovic05} and \cite{stverak09} found that the relative densities of the strahl and halo populations vary inversely with heliocentric distance, providing indirect evidence that the halo population may be formed from scattered strahl particles.}

A new analytical model for the strahl distribution was developed in \cite{horaites18}, by deriving exact solutions to the collisional kinetic equation in the asymptotic high-energy regime relevant to the strahl. This model was fit to eVDF data measured by the {\it Wind} satellite's SWE strahl detector measured at 1 AU, with remarkable agreement. In particular, it was shown that the model can predict how the angular width of this population scales with particle energy and background density. However, matching this model to 1 AU data also leads to the prediction that the strahl amplitude should {\it increase} with heliocentric distance---in contradiction with the trend observed by \cite{stverak09}. The model also predicts, in accordance with \cite{lemonsfeldman83}, that at a given energy the strahl should become narrower with heliocentric distance---the opposite radial trend that was reported in \cite{hammond96, graham17} in the outer heliosphere. Despite the successes of collisional model proposed in \cite{horaites18}, it may require improvement through the inclusion of additional physical processes. %e.g. kinetic instabilities

Examples of such processes are the kinetic instabilities that can be triggered by a non-Maxwellian electron distribution function. In this work we revisit the question of kinetic instabilites in the solar wind plasma, for an eVDF that is composed of parallel-drifting core and strahl components. For the first time we conduct the stability analysis using a realistic distribution function \citep[described above, from ][]{horaites18} for the strahl component, derived from the electron kinetic equation. 
%This model was introduced in \cite{horaites18} and was shown to successfully match the eVDF data at 1 AU.
 The strahl distribution function is rather nontrivial and, therefore, its stability analysis has to be conducted numerically. For this purpose, we employ the LEOPARD solver \citep{astfalk17}, which has been recently developed to analyze the stability of arbitrary gyrotropic distributions in a magnetized plasma.

We search for unstable modes that one may expect to be most relevant to a core-strahl eVDF: the Alfv\'en, magnetosonic, and whistler modes. These modes have been shown to be unstable for model distributions composed of drifting core and halo components. For our core-strahl model, we indeed find that the distributions are unstable to the Alfv\'en and magnetosonic modes with wavenumbers $k d_i\lesssim 1$. \red{The ion inertial length $d_i$ is defined:}

\begin{equation}\label{d_i_eq}
\red{d_i = v_A / \Omega_i,}
\end{equation}

\noindent \red{where $\Omega_i \equiv e B/m_p$ is the ion cyclotron frequency and $v_A$ is the Alfv\'en speed. } However, we have not been able to identify the whistler modes that would be directly excited by a Cherenkov resonance with the strahl particles; such modes appear to be damped overall. We therefore propose that for the strahl particles to be scattered by the whistler modes \citep[as in, e.g.][]{vocks05, saitogary07}, such modes should be generated not by the strahl electrons but rather transferred to small scales as a result of a turbulent cascade, say originating from the core-drift instabilities at~$kd_i\lesssim 1$. This result \red{provides a source for whistler waves that} is different and complementary to the previously discussed mechanism of whistler instabilities~\cite[e.g.,][]{forslund70, gary75, saeed17}. 

\red{The effect of whistler waves on the eVDF has been investigated in Particle-in-Cell (PIC) simulations. For instance, \cite{saitogary07b} demonstrated that a broadband spectrum of parallel propagating whistlers could significantly broaden the strahl population. We note though that in PIC simulations of whistler turbulence, e.g., \cite{saito08} and \cite{chang13}, the parallel temperature of the eVDF increases with time more quickly than the perpendicular temperature. This calls into question the effectiveness of whistler turbulence as a source of electron pitch-angle (perpendicular) scattering; at least insofar as one might try to explain the presence of the nearly-isotropic halo through direct strahl scattering, as suggested by \cite{stverak09}. However, these turbulent simulations did not include a strahl component, so the impact of the turbulence on this component has not to our knowledge been directly studied.} 

\red{We note that our model core-strahl eVDF omits the halo component; this choice is based on practical and theoretical considerations. First of all, halo-driven instabilities are already a well-studied topic that need not be rehashed here, as it is well known for instance that halo temperature anisotropy and core-halo drift can lead to linear instabilities. Furthermore, the inclusion of a halo component would complicate our model by introducing additional free parameters, such as halo anisotropy and relative density, which would needlessly expand the parameter space we wish to explore. We might reasonably try to simplify such a core-halo-strahl model by assuming a strictly isotropic (and monotonic with energy) halo, which would be consistent with the average properties of this component \citep{pierrard16}. But isotropic, monotonic distributions are always linearly stable \citep[e.g., ][]{clemmowdougherty69}, so the inclusion of a tenuous isotropic halo can be expected to have only a slight stabilizing effect, and should not introduce any new instabilities beyond those found with the more straightforward core-strahl model. For empirical support of this claim, see Appendix \ref{appendix}, in which we present a preliminary stability analysis of a fiducial core-halo-strahl eVDF. }

\section{Core-Strahl Distribution Function}
In order to conduct a stability analysis, we will model the electron velocity distribution function $f(\mu, v)$ as a core-strahl system. Since the distributions are assumed to be gyrotropic, we use as independent variables the velocity magnitude $v$, and cosine of the pitch angle $\mu$:
%, where  $\mu\equiv \bvec{\hat B} \cdot \bvec{v}/v = \cos \theta$, and $\xi \equiv (v/v_{th})^2$.
\begin{equation}
 \mu\equiv \bvec{\hat B} \cdot \bvec{v}/v,
\end{equation}
where the unit vector $\bvec{\hat B}$ points along the (Parker spiral) magnetic field, in the anti-sunward orientation. 

Let us designate our core and strahl model functions as $f_c(\mu, v)$, $f_s(\mu, v)$, respectively. The total distribution is then $f = f_c + f_s$, and the total density $n$ is given by: 
\begin{equation}
\int f(\bvec{v}) d^3v = 2\pi \int_0^\infty \int_{-1}^1 f(\mu, v) v^2 d\mu dv = n. 
\end{equation}
The core Maxwellian distribution $f_c(\mu, v)$ is allowed to drift (sunward) at parallel velocity $v_d$ relative to the protons. This distribution has the form:
\begin{equation}\label{fc_eq}
f_c(\mu, v) = \frac{n_c}{\pi^{3/2} v_{th}^3} \exp \left( \frac{-v^2 + 2 \mu v v_d - v_d^2}{v_{th}^2}\right),
\end{equation}
where $n_c$, $v_{th}$ represent the electron core density and thermal speed, respectively. 

Our model for the strahl distribution, $f_s(\mu, v)$, comes from \cite{horaites18}, Eq.~(38):
\begin{equation}\label{fs_eq}
f_s(\mu, v) =  C_0 A(v) \frac{n_c}{v_{th}^3} \left(\frac{v}{v_{th}}\right)^{2\epsilon} \exp\left[\tilde{\gamma}\Omega (v/v_{th})^4 (1-\mu) \right].
\end{equation}
\noindent In this expression we use the ``effective'' Knudsen number introduced in \cite[][Eq.~26]{horaites18},
\begin{eqnarray}\label{gamma_eq}
%{\tilde \gamma}(x)=T^2/\left(2\pi e^2 \Lambda n x\right),
{\tilde \gamma}(x)=T^2/\left(2\pi e^2 \Lambda n_c x\right),
\end{eqnarray}
where $T(x)=m_e v_{th}^2/2$ is the core electron temperature, 
%$n(x)\approx n_c$~is the density, 
$\Lambda$~is the Coulomb logarithm, and $x$~is the heliospheric distance.  We let the parameters $\epsilon$ and $\Omega$ be given by empirical measurements \citep{horaites18}, representative of the typical ($\tilde{\gamma} = 0.75$) fast wind: $\epsilon \equiv -2.14$, $\Omega \equiv -0.3$. A summary of these constants, that do not vary throughout our analysis, are given in Table \ref{constant_table}. The analytic derivation of the strahl shape~(\ref{fs_eq}) does not however allow one to obtain the overall strahl amplitude. In \cite[][]{horaites18} the constant~$C_0$ in the strahl distribution was therefore estimated from matching with the observational data. In our current consideration it is kept as a free parameter, and the stability analysis is performed for a range of possible strahl amplitudes~$C_0$. 

In Eq.~(\ref{fs_eq}) we also introduced a truncation function $A(v)$, to ensure that $f_s \rightarrow 0$ as $ v \rightarrow 0$:
\begin{equation}
A(v) =  \frac{1}{1 + a (v/v_{th})^b},
\end{equation}
\noindent where we defined constants $a\equiv 10$, $b \equiv 2 \epsilon -4$. The form of this low-energy truncation function is somewhat arbitrary, but its introduction is necessary since Eq.~(\ref{fs_eq}) was derived assuming $(v/v_{th})^2 \gg 1$. The function $A(v)$ artificially modifies the strahl only at $v<v_{th}$, where the distribution function is anyway dominated by the core component.  
   
\begin{table}
\centering
\begin{tabular}{| c || c |}
\hline
$\tilde{\gamma}$ & 0.75\\
$\Omega$ & -0.3 \\
$\epsilon$ & -2.14 \\
a & 10 \\
b & 2 $\epsilon$ - 4 = -8.28\\
\hline
\end{tabular}
\caption{Constants used in Eq.~(\ref{fs_eq}), that are not altered throughout our analysis.}
\label{constant_table}
\end{table}

%\begin{equation}
%\begin{split}
%&\epsilon \equiv -2.14,\\
%&\Omega \equiv -0.3,
%\end{split}
%\end{equation}

As an input to the kinetic solver, we will assume a steady state where the parallel current $J_\parallel$ is zero. That is, we require:
\begin{equation}
%\re
J_\parallel \equiv \int v_\parallel f(\bvec{v}) d^3v =  2\pi \int_0^\infty \int_{-1}^1 f(\mu, v) v^3 \mu d\mu dv= 0.
\end{equation} 
We can decompose total parallel current into contributions from the core and strahl, that is:
\begin{equation}
J_\parallel = J_{\parallel,c} + J_{\parallel,s}.
\end{equation}

The analytic form of the core contribution $J_{\parallel, c}$ follows from Eq.~(\ref{fc_eq}):
\begin{equation}
J_{\parallel,c} = n_c v_d,
\end{equation}
and we can therefore write a simple expression for the core drift $v_d$ that ensures $J_\parallel = 0$:
\begin{equation}\label{vd_eq}
v_d = -J_{\parallel,s}/n_c.
\end{equation}
Eq.~(\ref{vd_eq}) suggests a simple procedure for finding the core drift $v_d$ that ensures $J_\parallel = 0$. First, the strahl distribution $f_s(\mu, v)$ is integrated numerically to find $J_{\parallel,s}$, and this value is substituted into Eq.~(\ref{vd_eq}) to find $v_d$ ($n_c$ is given). 

In the analysis presented in section \ref{analysis_sec}, we will assume a set of plasma parameters that are representative of the fast wind at 1 AU, that we will use as a baseline. These fiducial plasma parameters---the core density $n_c$, core temperature $T$, magnetic field strength $B$---are presented in table \ref{phys_param_table}. These parameters are consistent with a $\tilde{\gamma} = 0.75$, $\beta_e = 0.307$ plasma, where $\tilde{\gamma}$ is defined in Eq.~(\ref{gamma_eq}) and the electron beta, $\beta_e$, is defined:
\begin{equation}\label{beta_eq}
\beta_e \equiv {8 \pi n_c T}/{B^2}.
\end{equation}

\noindent The strahl amplitude $C_0$, typical of the $\tilde{\gamma} = 0.75$ solar wind \citep{horaites18}, is also presented in the table.

\begin{table}
\centering
\begin{tabular}{| c || c |}
\hline
$n_c$ ($\approx n$) & 4 cm$^{-3}$\\
T & 12.21 eV \\    % = sqrt(\tilde{\gamma} * n / 0.02012)
B & 8 nT \\
\hline
$C_0$ & 0.234\\
\hline
\end{tabular}
\caption{The set of physical constants presented in this table are used as a baseline for our model. These constants are consistent with a $\beta_e = 0.307$, $\tilde{\gamma} = 0.75$ plasma. When we investigate the effect of $\beta_e$ on stability, a different set of $n_c$, $T$, $B$ will be used (see section \ref{analysis_sec}). The strahl amplitude $C_0$ reported here is a typical value that was measured in the $\tilde{\gamma} = 0.75$ fast wind \citep{horaites18}. The parameter $C_0$ will be varied to investigate its effect on stability, to produce Figs.~\ref{C_scan_kaw_plot} and \ref{C_scan_ms_plot}.}
\label{phys_param_table}
\end{table}

\section{Stability Analysis}\label{analysis_sec}

%If the distribution is found to be unstable, we may expect the growth rate of the instability to depend on the amplitude of the strahl relative to the core. In the table \ref{table_vd}, we vary $C_0$ for fixed $\tilde{\gamma}$ and calculate (numerically) the drift velocity $v_d$ (normalized to the thermal speed $v_{th}$) required to ensure $J_\parallel=0$. 

We analyze the linear stability of the eVDF using the LEOPARD solver \citep{astfalk17}, which can calculate the dispersion relation $\omega(\bvec{k})$ for an arbitrary gyrotropic electron distribution. The imaginary part of our solutions determines the stability of the particular wavemode: stable for $Im(\omega)\leq 0$  and unstable for $Im(\omega)>0$. We will assume that the large-scale variation in any plasma parameters (density, temperature, E- and B-fields) is slow enough that it can be neglected on the spatial scale of the waves. For simplicity, we assume the  background electric field is zero. 

The final solution $\omega(\bvec{k})$ is found through an iterative scheme, that converges most efficiently when the initial guesses for $\omega$, $\bvec{k}$ are near an actual solution. As a starting point for our analysis, we use fully kinetic dispersion relations of the kinetic Alfv\'en, fast magnetosonic, and whistler branches in a Maxwellian plasma given in, e.g., \cite{told16}. By smoothly varying the parameters in our model, we can then scan through different propagation angles and parameter regimes to explore the stability of the branches with respect to our core-strahl distribution.

%\begin{equation}\label{whistler_analytic_eq}
%\frac{k^2 c^2}{\omega_{pe}^2}\approx \frac{\omega^2}{\omega(\Omega_e \cos \theta-\omega)},
%\end{equation}

%\noindent where $\theta$ denotes the angle between the magnetic field $\bvec{B}$ and the wavevector $\bvec{k}$. The imaginary part $\gamma\equiv Im(\omega)$ of our solutions $\omega(\bvec{k})$ determines the stability of the particular wavemode: stable for $\gamma<0$  and unstable for $\gamma>0$.

For each branch, we scanned through the propagation angles $0^\circ < \theta < 89^\circ$, with $1^\circ$ resolution. At these angles, we varied the strahl amplitude across the values $C_0 = 0.000, 0.050, 0.100, 0.200$ while holding $\beta_e = 0.307$ fixed. Holding $C_0 = 0.234$ fixed and scanning through these same angles, we varied beta across the values $\beta_e = 0.307, 0.500, 0.700$. We completed these scans for wavenumbers $0.1 < k d_i < 1.0$ for the magnetosonic mode and $0.1 < k d_i < 4.0$ for the KAW mode. We conducted a similar analysis over the range of wavenumbers $1 < k d_i < 40$ for the whistler mode; however, we investigated only angles $0< \theta <79^\circ$ and $\theta = 180^\circ$.

\begin{figure}
\includegraphics[width=1\linewidth]{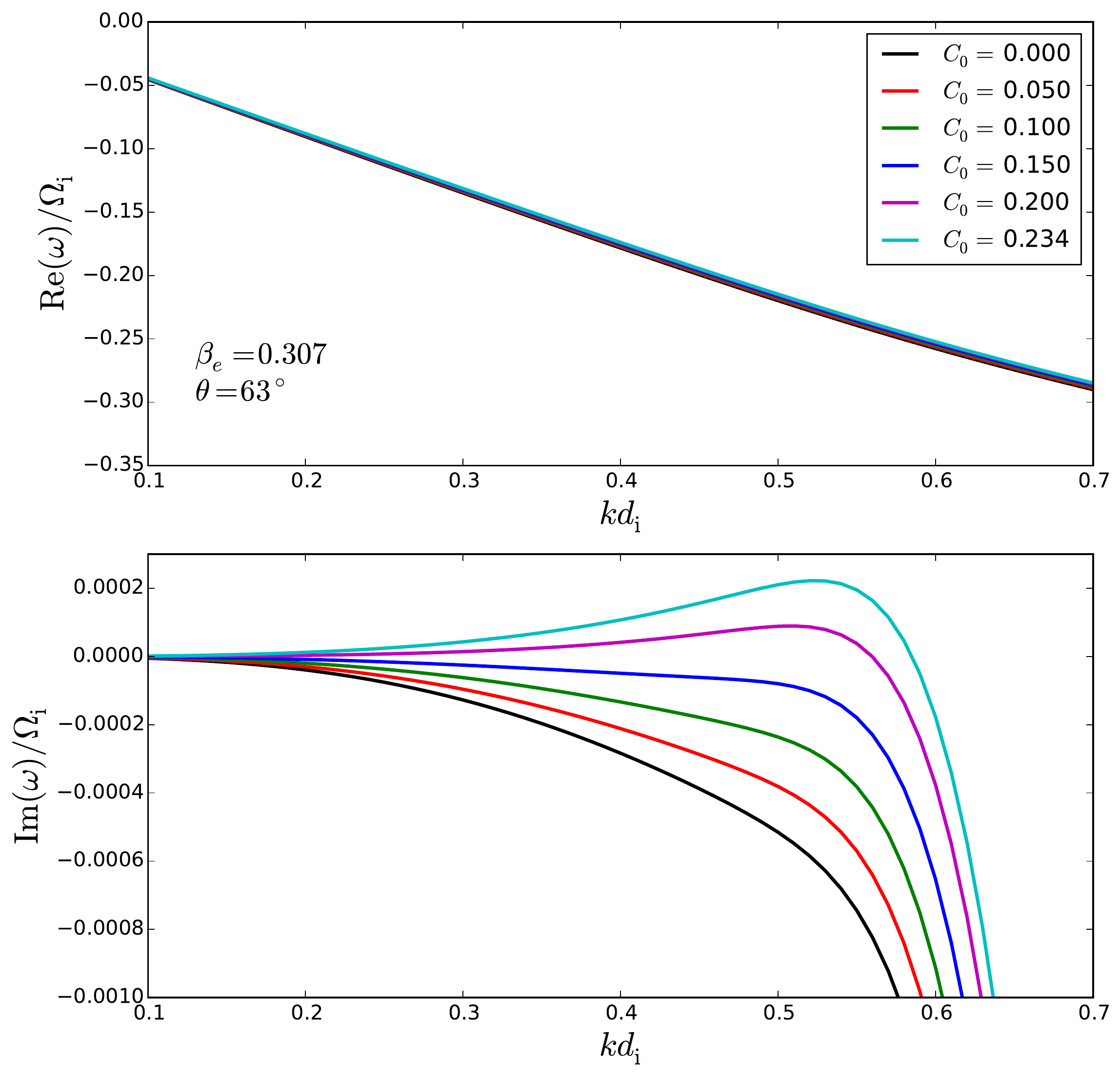}
\caption{{\it KAW---Less oblique}. Real (upper) and imaginary (lower) parts of the KAW dispersion relation, shown for different strahl amplitudes $C_0$. We set $\theta=63^\circ$ and $\beta_e=0.307$ for all calculations, but vary the strahl amplitude $C_0$ (and the core drift $v_d$, by Eq.~(\ref{vd_eq})). The distribution becomes unstable for $C_0 \gtrsim 0.20$.}
\label{C_scan_kaw_plot}
\end{figure}

In Fig.~\ref{C_scan_kaw_plot}, we show our numerical results for the kinetic Alfv\'en (KAW) mode in the range of wavenumbers $ 0 \lesssim k d_i \lesssim 0.7$.
%      In all figures, frequencies and growth rates are normalized to the ion cyclotron frequency $\Omega_i \equiv e B/m_p$ and wavenumbers are normalized to the ion inertial length (\ref{d_i_eq}).
     \red{ In all figures, frequencies and growth rates are normalized to the ion cyclotron frequency $\Omega_i$ and wavenumbers are normalized to the ion inertial length $d_i$ (Eq.~\ref{d_i_eq}). }
 We solve for the dispersion relation for different strahl amplitudes $C_0$ (Eq. \ref{fs_eq}), represented as different lines in Fig.~\ref{C_scan_kaw_plot}. Note that adjusting $C_0$ also requires adjustment of the core drift $v_d$ according to Eq.~(\ref{vd_eq}). All other physical parameters and constants are as listed in tables \ref{constant_table} and \ref{phys_param_table}. The propagation angle is set to $\theta = 63^\circ$. We see that the distribution becomes unstable if the strahl amplitude is sufficiently large, i.e. in the regime $C_0 \gtrsim 0.20$. We also note that since $Re(\omega<0$), the waves propagate with a {\it sunward} parallel phase speed. 

\begin{figure}
\includegraphics[width=1\linewidth]{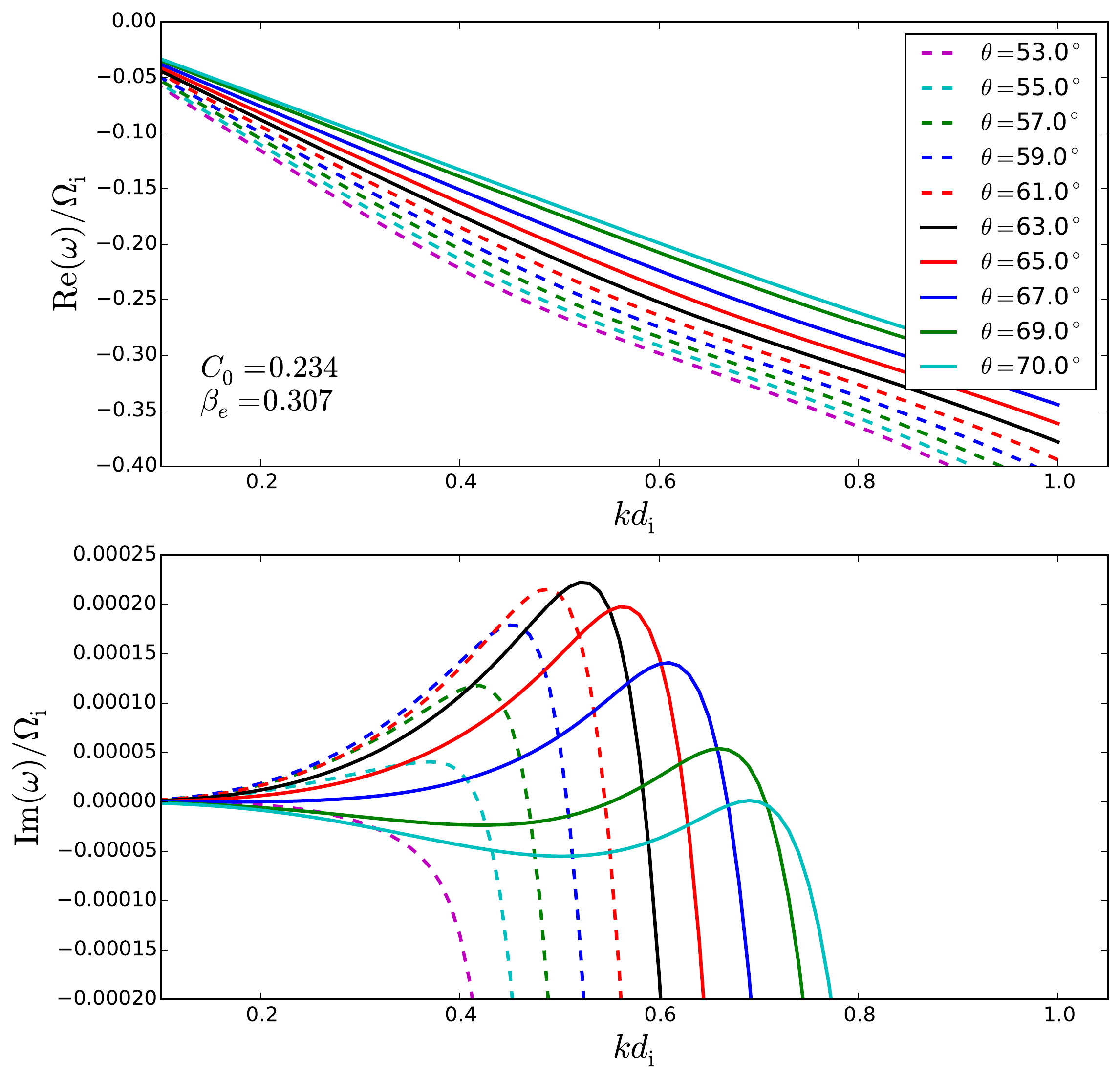}
\caption{{\it KAW---Less oblique}. Real (upper) and imaginary (lower) parts of the KAW dispersion relation, shown for different propagation angles $\theta$. We set $\beta_e=0.307$ and $C_0=0.234$ for all calculations, but vary the propagation angle $\theta$. The distribution is unstable in the range $55^\circ \lesssim \theta \lesssim 69^\circ$, and is maximally unstable at $\theta \approx 63^\circ$.}
\label{theta_scan_kaw_plot}
\end{figure}

The growth rate of the KAW instability appears particularly sensitive to the propagation angle $\theta$; see Fig.~\ref{theta_scan_kaw_plot}. Here we hold the strahl amplitude $C_0$ constant, and instead vary the propagation angle $\theta$. We see the distribution is only unstable in a range of moderately oblique angles $55^\circ \lesssim \theta \lesssim 69^\circ$, and is maximally unstable at $\theta \approx 63^\circ$. 

\begin{figure}
\includegraphics[width=1\linewidth]{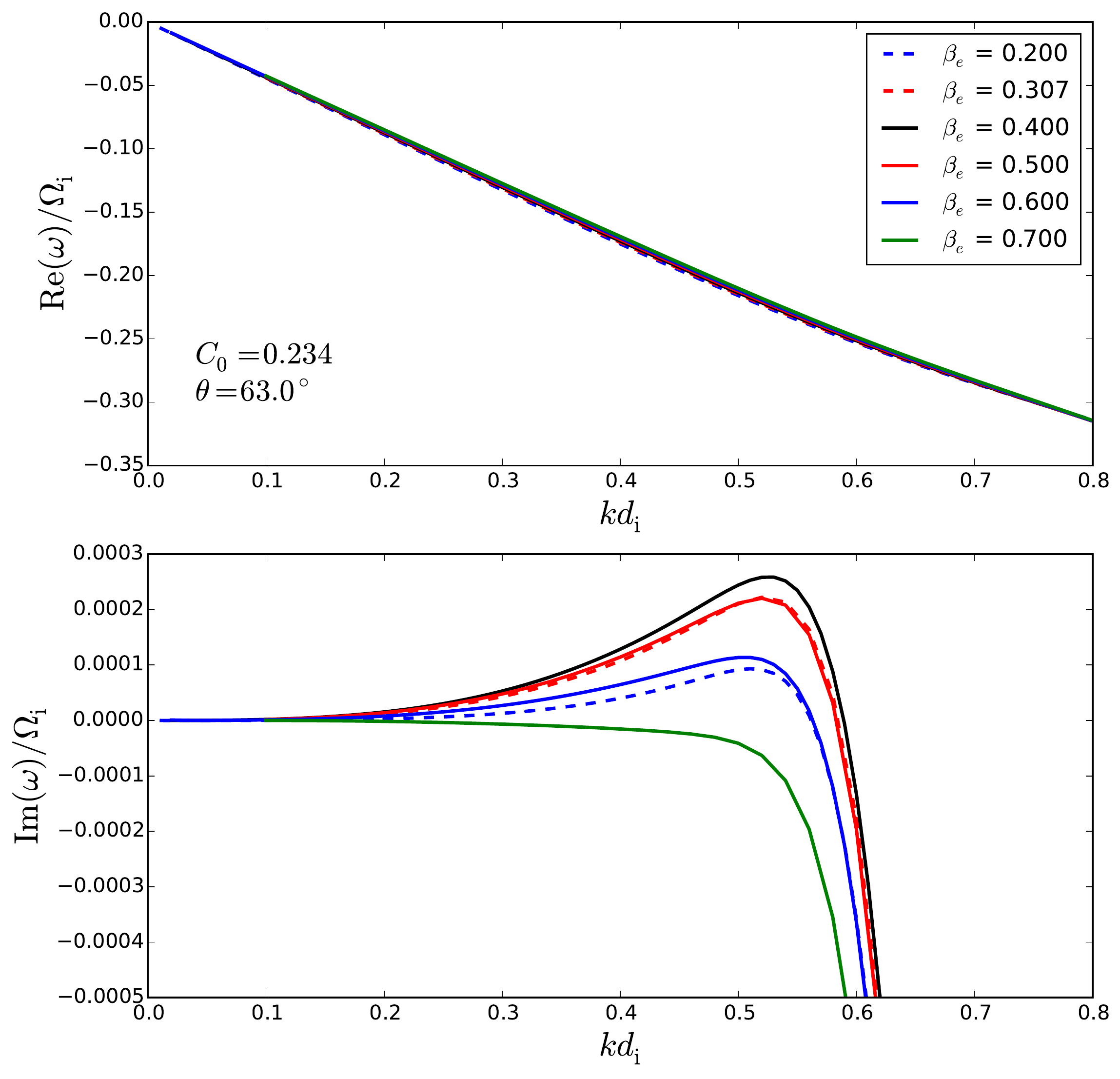}
\caption{{\it KAW---Less oblique}. Real (upper) and imaginary (lower) parts of the KAW dispersion relation, shown for different electron betas ($\beta_e$). We set $\theta=63^\circ$ and $C_0=0.234$ for all calculations, but vary the electron beta $\beta_e$. The distribution is unstable in the range $0.2 \lesssim \beta_e \lesssim 0.6$, and is maximally unstable at $\beta_e \approx 0.4$.}
\label{beta_scan_kaw_plot}
\end{figure}

In Fig.~\ref{beta_scan_kaw_plot}, we investigate the KAW instability's dependence on the electron beta. Each line in the figure shows the dispersion relation for a different $\beta_e$. The line corresponding with $\beta_e=0.307$ is generated using the physical parameters given in Table \ref{phys_param_table}. The other lines, representing different $\beta_e$, require at least one of the parameters $n_c$, $T$, $B$ to be adjusted. Let the fiducial parameters listed in table \ref{phys_param_table} be written as $n_{c,0}$, $T_0$, $B_0$, and let $\beta_{e,0}\equiv 0.307$. To investigate a different plasma beta, $\beta_e = \alpha \beta_{e,0}$ (where $\alpha \neq 1$), we choose $n_c$, $T$, $B$ in the following manner:
%\begin{subequations}
%\begin{centering}
 \begin{align}
& T = \alpha T_0, \label{beta_var_eq1} \\
& n_c = \alpha^2 n_{c,0}, \label{beta_var_eq2}\\
& B = \alpha B \label{beta_var_eq3}.
 \end{align}
%\end{centering}
%\end{subequations}

%\begin{eqnarray}\label{beta_var_eq}
%& T = \alpha T_0, \\
%& n_c = \alpha^2 n_{c,0},\\
%& B = \alpha B.
%\end{eqnarray}

This scheme allows us to scale $\beta_e$ while holding $\tilde{\gamma}$ and the Alfv\'en speed $v_A$ constant. Holding $\tilde{\gamma}$ constant enables ready comparison with \cite{horaites18}, while holding $v_A$ constant follows the precedent set by \cite{gary94}. We see in Fig.~\ref{beta_scan_kaw_plot} that the distribution is unstable in the range $0.2 \lesssim \beta_e \lesssim 0.6$, and is maximally unstable at $\beta_e \approx 0.4$.

\begin{figure}
\includegraphics[width=1\linewidth]{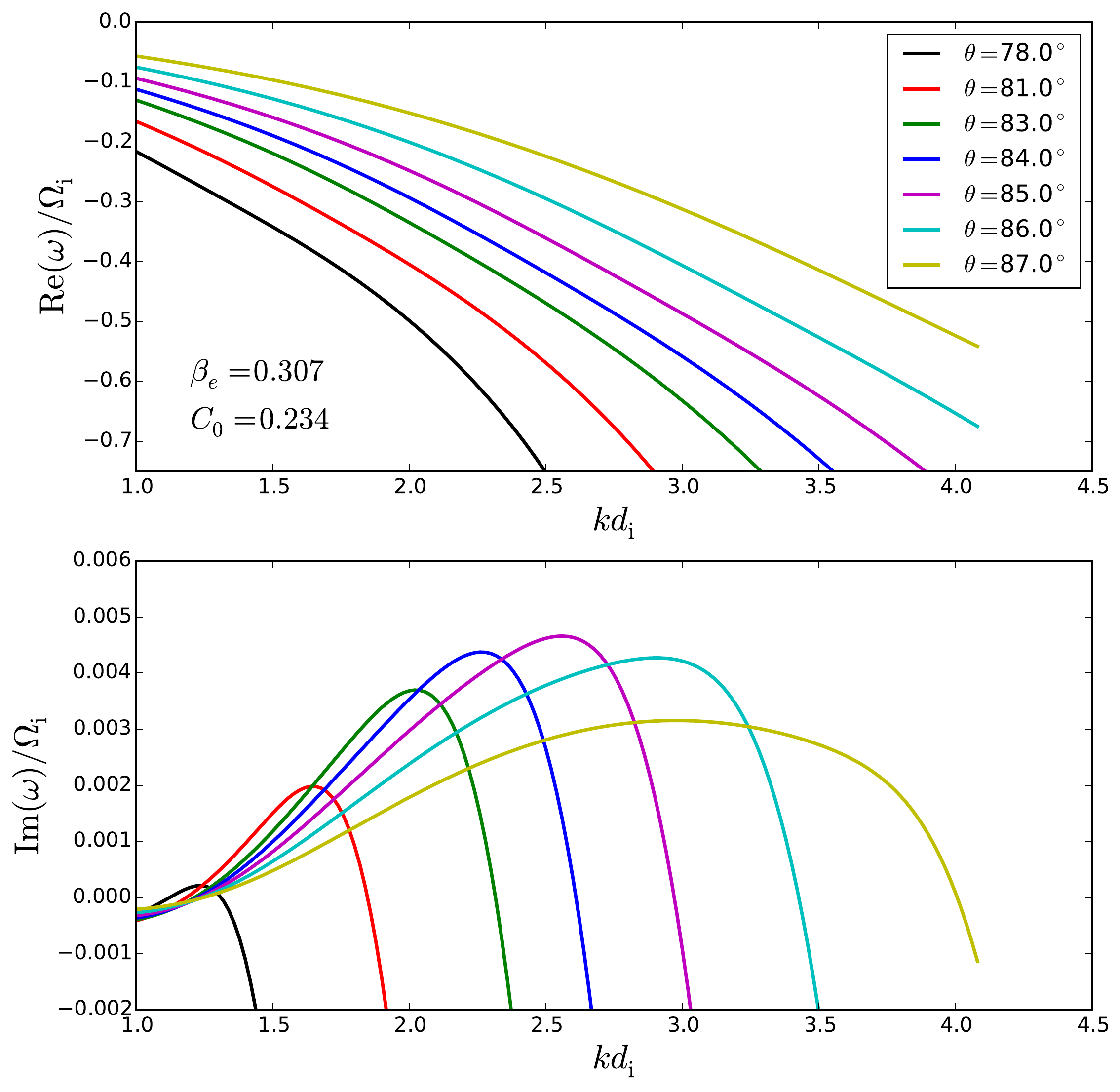}
\caption{{\it KAW---More oblique}. Real (upper) and imaginary (lower) parts of the KAW dispersion relation, shown for different propagation angles $\theta$. We set $\beta_e=0.307$ and $C_0=0.234$ for all calculations, but vary the propagation angle $\theta$. Here we see an instability in the range of wavenumbers $1 \lesssim k d_i \lesssim 4$. The distribution is unstable in the range $78^\circ \lesssim \theta \lesssim 87^\circ$, and is maximally unstable at $\theta \approx 85^\circ$.}
\label{theta_scan_kaw_oblique_plot}
\end{figure}

The distribution also exhibits a second KAW instability; the unstable regime falls in the range of wavenumbers $1 \lesssim k d_i \lesssim 4$, see figure \ref{theta_scan_kaw_oblique_plot}. The unstable waves are more oblique here, with propagation angles falling in the range $78^\circ \lesssim \theta \lesssim 87^\circ$. Although the waves here possess growth rates that are about an order of magnitude larger than the growth rates of the less oblique KAW waves (figure \ref{theta_scan_kaw_plot}), their obliquity makes them less able to couple to whistler waves, and therefore these waves are less relevant to generating whistler turbulence that may scatter the strahl (see section \ref{discussion}).

\begin{figure}
\includegraphics[width=1\linewidth]{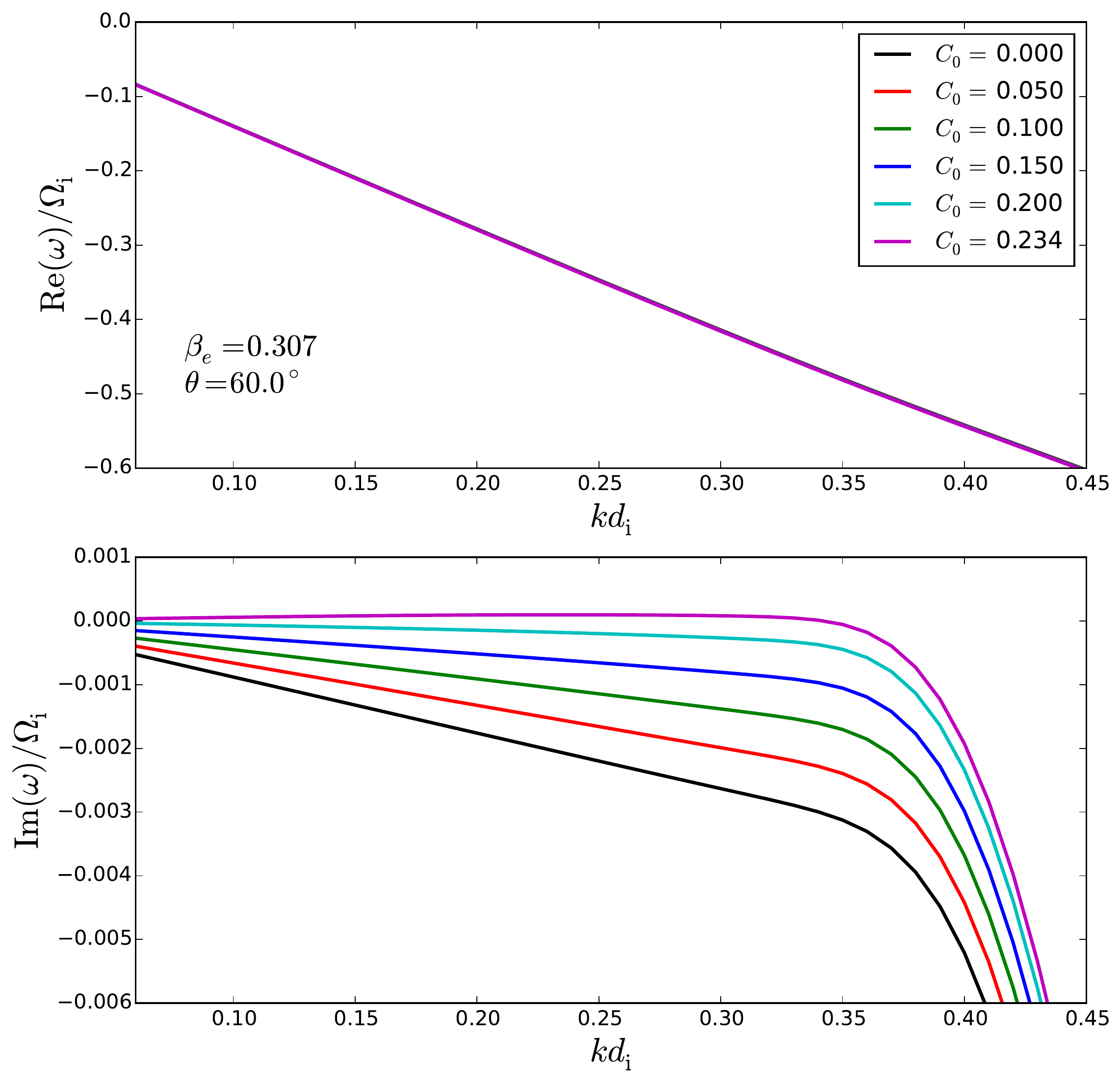}
\caption{{\it Magnesotonic}. Real (upper) and imaginary (lower) parts of the magnetosonic dispersion relation, shown for different strahl amplitudes $C_0$. We set $\theta=60^\circ$ and $\beta_e=0.307$ for all calculations, but vary the strahl amplitude $C_0$ (and the core drift $v_d$, by Eq.~(\ref{vd_eq})). The distribution becomes unstable for $C_0 \gtrsim 0.20$.}
\label{C_scan_ms_plot}
\end{figure}

\begin{figure}
\includegraphics[width=1\linewidth]{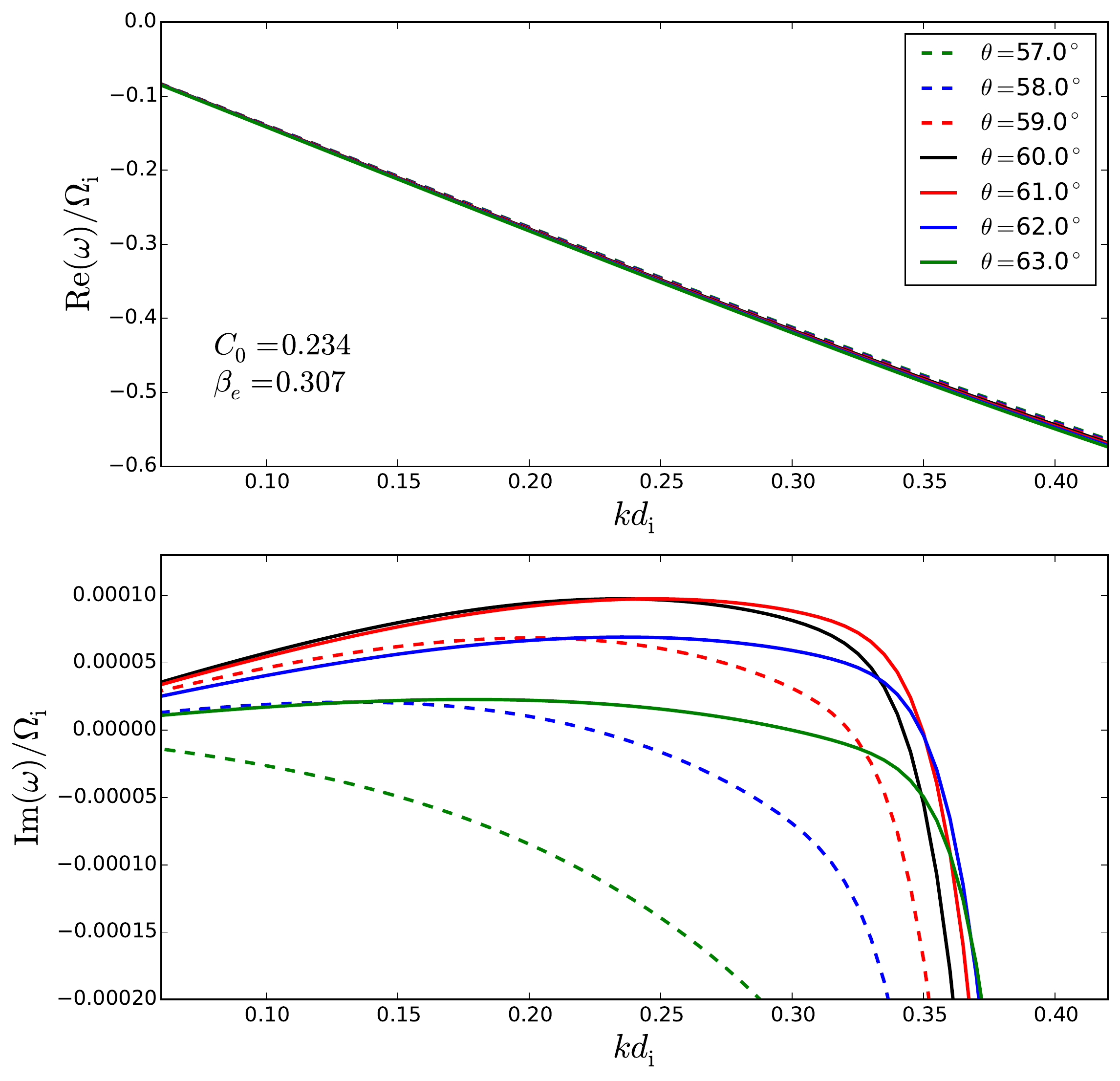}
\caption{{\it Magnetosonic}. Real (upper) and imaginary (lower) parts of the magnetosonic dispersion relation, shown for different electron betas ($\beta_e$). We set $\beta_e=0.307$ and $C_0=0.234$ for all calculations, but vary the propagation angle $\theta$. The distribution is unstable in the range $58^\circ \lesssim \theta \lesssim 63^\circ$, and is maximally unstable at $\theta \approx 60^\circ$.}
\label{theta_scan_ms_plot}
\end{figure}

\begin{figure}
\includegraphics[width=1\linewidth]{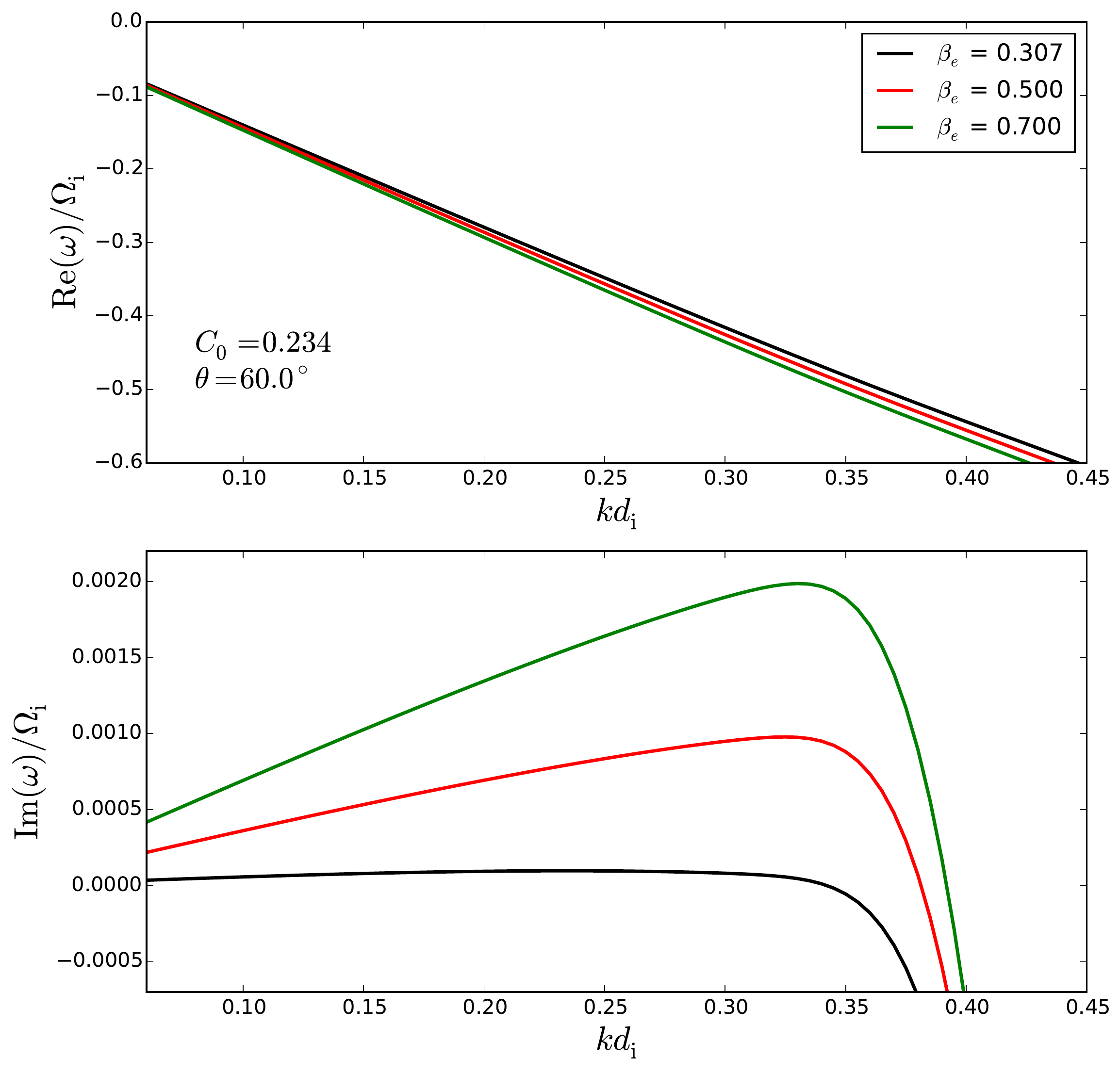}
\caption{{\it Magnetosonic}. Real (upper) and imaginary (lower) parts of the magnetosonic dispersion relation, shown for different electron betas $\beta_e$. We set $\theta=60^\circ$ and $C_0=0.234$ for all calculations, but vary the electron beta $\beta_e$. The distribution is unstable in the range $0.3 \lesssim \beta_e \lesssim 0.7$, \red{with monotonically increasing growth as $\beta_e$ increases in this range.}}
\label{beta_scan_ms_plot}
\end{figure}

Our core-strahl model function is also unstable to the magnetosonic mode, as shown in Figs.~\ref{C_scan_ms_plot}, \ref{theta_scan_ms_plot}, \ref{beta_scan_ms_plot}. These plots are analogous to Figs.~\ref{C_scan_kaw_plot}, \ref{theta_scan_kaw_plot}, \ref{beta_scan_kaw_plot} respectively, but for the magnetosonic mode instead of the KAW mode. In figure \ref{C_scan_ms_plot}, we see that the function is unstable for the baseline parameters shown in Table \ref{phys_param_table}. We see in Fig.~\ref{C_scan_ms_plot} that for the magnetosonic mode, as with the KAW mode, the distribution is unstable when the strahl amplitude is sufficiently large, i.e. in the regime $C_0 \gtrsim 0.20$. Fig.~\ref{theta_scan_ms_plot} demonstrates that the largest growth rate occurs at $\theta\approx 60^\circ$, a moderately oblique angle. As with the KAW mode, the instability occurs for $Re(\omega)<0$; i.e., the waves propagate toward the sun. Fig.~\ref{beta_scan_ms_plot} displays the instability's dependence on $\beta_e$; here we see that the distribution is unstable in the range $0.3 \lesssim \beta_e \lesssim 0.7$, \red{with monotonically increasing growth as $\beta_e$ increases in this range.} The parameter $\beta_e$ was adjusted by varying $n_c$, $T$, $B$ in the scheme described by Eqs.~(\ref{beta_var_eq1}-\ref{beta_var_eq3}) and the preceding paragraph.

Finally, we note that no instabilities were found for the whistler mode. Although we did observe a whistler mode, all solutions were damped ($Im(\omega)<0$). In addition to searching through the angles $0^\circ$-$79^\circ$, we checked the stability of antiparallel-propagating ($\theta = 180^\circ$) waves as well. We note that the whistler waves observed recently observed by \cite{stansby16}, which were attributed to the presence of a heat flux instability, exhibited only parallel and anti-parallel propagation angles.
%We searched for parallel- (all angles $\theta = 0^\circ-89^\circ$, with $1^\circ$ spacing) and antiparallel-propagating ($\theta = 180^\circ$) waves, in the range of wavenumbers $1 \lesssim k d_i \lesssim 40$. 
For brevity, we do not show a plot of the whistler results, but note only that the absence of a whistler instability is in contrast with the results of \cite{gary94}, \red{who use a core-halo electron model to show linear theory growth of the whistler heat flux instability}.

\section{Discussion and Conclusions}\label{discussion}

We here summarize our main results:

1. We analyzed the kinetic stability of a core-strahl eVDF, using a realistic distribution function that is representative of the typical fast solar wind at 1 AU. In agreement with \cite{gary75}, we observe the eVDF is unstable to the magnetosonic and KAW modes in the range of wavenumbers $k d_i \lesssim 1$. These modes are driven by the Landau resonance with the core electrons, i.e. at parallel velocities $v_\parallel = Re(\omega)/k_\parallel \sim (-v_A)$. The negative sign here indicates that the resonant electrons are traveling slightly sunward, relative to the proton bulk flow. In fact, this resonant velocity falls somewhere between the peaks of the electron and ion VDFs. These results may be fairly insensitive to the particular model used for the core distribution---we note that \cite{forslund70} found similar waves and resonances for a core distribution that was non-Maxwellian, i.e. a Maxwellian distorted by a Spitzer-H\"arm electric field \citep{spitzerharm53}.

2. \red{Our linear theory analysis does not yield a whistler instability.} This begs comparison with the results of \cite{gary94}, in which a whistler heat flux instability could be excited by a skewed non-thermal eVDF. The difference is likely due to the fact that our model describes a core-strahl rather than a core-halo distribution. That is, the suprathermal heat-flux carrying electrons are modeled by Eq.~(\ref{fs_eq}), rather than by a drifting Maxwellian (or other function that is isotropic in its own frame). 
%Our results suggest that relating a whistler instability to a heat flux may be misleading. 
%Indeed, in our case a heat flux is associated with a large-amplitude strahl, rather than a drifting halo. Such a situation is common in the cases when a large-amplitude strahl is present~\citep[e.g., ][]{pilipp87}. Nevertheless, as we have demonstrated, the heat-carrying strahl electrons do not trigger the whistler instability. 

The cases where the direct observation of whistler waves has been associated with a large electron heat flux appear to be limited to the slow wind, where the strahl tends to be less prominent \cite[see figures 5 and 9, ][]{lacombe14}. We propose that further investigation of these events could reveal the presence of a drifting halo; however, to our knowledge such a study has not yet been undertaken. 

In the cases where the heat flux is mostly accounted for by the electron strahl, on the other hand, our present work suggests that a large electron heat flux does not necessarily trigger a whistler instability. As a large heat flux is typically associated with a prominent strahl~\citep[e.g., ][]{pilipp87}, caution must be used when parametrizing the stability of the eVDF in terms of the heat flux, so that the physics of core-halo and core-strahl distributions are not confounded.

3. Remarkably, the observed instability thresholds at $C_0\sim 0.2$ appear very close to the average strahl amplitude $C_0=0.234$ observed in \cite{horaites18}, for the $\tilde{\gamma}=0.75$ fast wind. This indicates that the strahl amplitude may be regulated by the instabilities, as has been suggested by other authors. We note that the peak growth rates of the instabilities detected here, with $Im(\omega) / \Omega_i > 10^{-4}$, are just fast enough to be relevant for the solar wind strahl. That is, a typical strahl particle traveling as speed $10^4$ km/sec at 1 AU, where the local ion cyclotron frequency is $\Omega_i \sim 1$ Hz, traverses a typical scale height ($\sim$1 AU) in a time $\sim 10^4/ \Omega_i$. However, since the observed magnetosonic and KAW modes do not resonate directly with the strahl, we must assume that they could only regulate the strahl amplitude by some indirect mechanism \citep[see also, ][]{tong18}. 
%We propose that the observed instabilities that are less oblique ($\theta \approx 60^\circ$) may generate whistler waves that are capable of scattering the strahl, via turbulent wave-wave coupling.
%Indeed, the presence of observed wave fluctuations at wavenumbers $k d_i \gtrsim 1$ are difficult to explain in terms of the turbulent cascade that isgenerated on much larger scales; this Alfv\'enic turbulence is expected to dissipate at $k d_i \sim 1$ ({\bf CITATION}).

When strong turbulence of moderately oblique kinetic Alfv\'en and magnetosonic modes is generated at scales $k d_i \lesssim 1$, it can produce whistler fluctuations at the scale $k d_i \sim 1$. These nonlinear fluctuations could then cascade into moderately oblique whistlers at even smaller scales, where they can interact with the strahl electrons, providing their \red{``anomalous''} scattering via the cyclotron resonance. This mechanism may also provide an explanation for the moderately oblique whistler-like fluctuations observed in the solar wind by~\cite{narita16}.

%The magnetosonic and whistler instabilities found here represent a new source of waves in the regime $k d_i \lesssim 1$. 

%\begin{figure}
%\includegraphics[width=1\linewidth]{plots/strahl_spectrum_9_19950130_phys.eps}
%\caption{\label{strahl_spectrum_fig} An example strahl spectrum $f_s$ (physical units), after applying an automated procedure for removing the halo population. This plot can be compared with figure 3, \citep{fitzenreiter98}, which shows the ``raw'' spectrum. The variables $\theta_S$, $\phi_S$ are spherical (GSE) coordinates that describe the velocity direction of the measured electrons. The angle $\theta_S=0$ corresponds with the ecliptic. Note that the detector's 12 anodes are not evenly spaced in $\theta_S$. The magnetic field direction $\bvec{\hat B}$ is determined through the process of nonlinear fitting described in section \ref{width_obs_sec}, and is shown in this example by a ``+'' symbol.}% with a significant gap between anodes near the $\theta_S=0$, so that the vertical axis of this plot is somewhat distorted.}
%\end{figure}

\appendix

\section{Comparison with a Core-Halo-Strahl Model}\label{appendix}

\red{ In the preceding analysis, we have ignored the halo component, which appears ubiquitously in solar wind eVDFs. Our reasons for omitting this component were described in section \ref{intro_sec}.  In particular, we argued that if we were to assume an isotropic halo, its presence would likely have only a slight stabilizing effect on the distribution. In order to test this claim, we here add a halo component to our model eVDF, and compare the dispersion relations derived from the core-halo-strahl function with those of the foregoing core-strahl model. Rather than conduct an in-depth analysis, we assume some representative halo function with fixed parameters, so as not to introduce any new free parameters to our model. }

\red{ We will represent the core and strahl distributions by Eqs.~\ref{fc_eq} and \ref{fs_eq}, as elsewhere in this paper. Let us then model the halo distribution $f_h(\mu, v)$ as an isotropic kappa function \cite[e.g., ][]{maksimovic05, stverak09}: }

\red{
\begin{equation}\label{fh_eq}
f_h(\mu, v) = A_h  \Big[ 1 + \frac{m_e v^2}{(2 \kappa - 3)T_h}\Big]^{(-\kappa-1)},
\end{equation}
}

\noindent \red{where $m_e$ is the electron mass, and $A_h$ is defined:}

\red{
\begin{equation}\label{ah_eq}
A_h = n_h  \Big[ \frac{m_e}{\pi (2 \kappa - 3) T_h } \Big]^{(3/2)} \frac{\Gamma(\kappa + 1)}{ \Gamma(\kappa - 1/2) },
\end{equation}
}

%Eq.~\ref{fh_eq} describes an isotropic distribution, and there are no drift or temperature anisotropy terms that have been associated with instabilities. 

\noindent \red{ 
where $\Gamma()$ is the gamma function. Table \ref{halo_param_table} lists our fiducial values for the density $n_h$, temperature $T_h$, and kappa parameter $\kappa$ that appear in Eqs.~\ref{fh_eq} and \ref{ah_eq}.}

\begin{figure}
\includegraphics[width=1\linewidth]{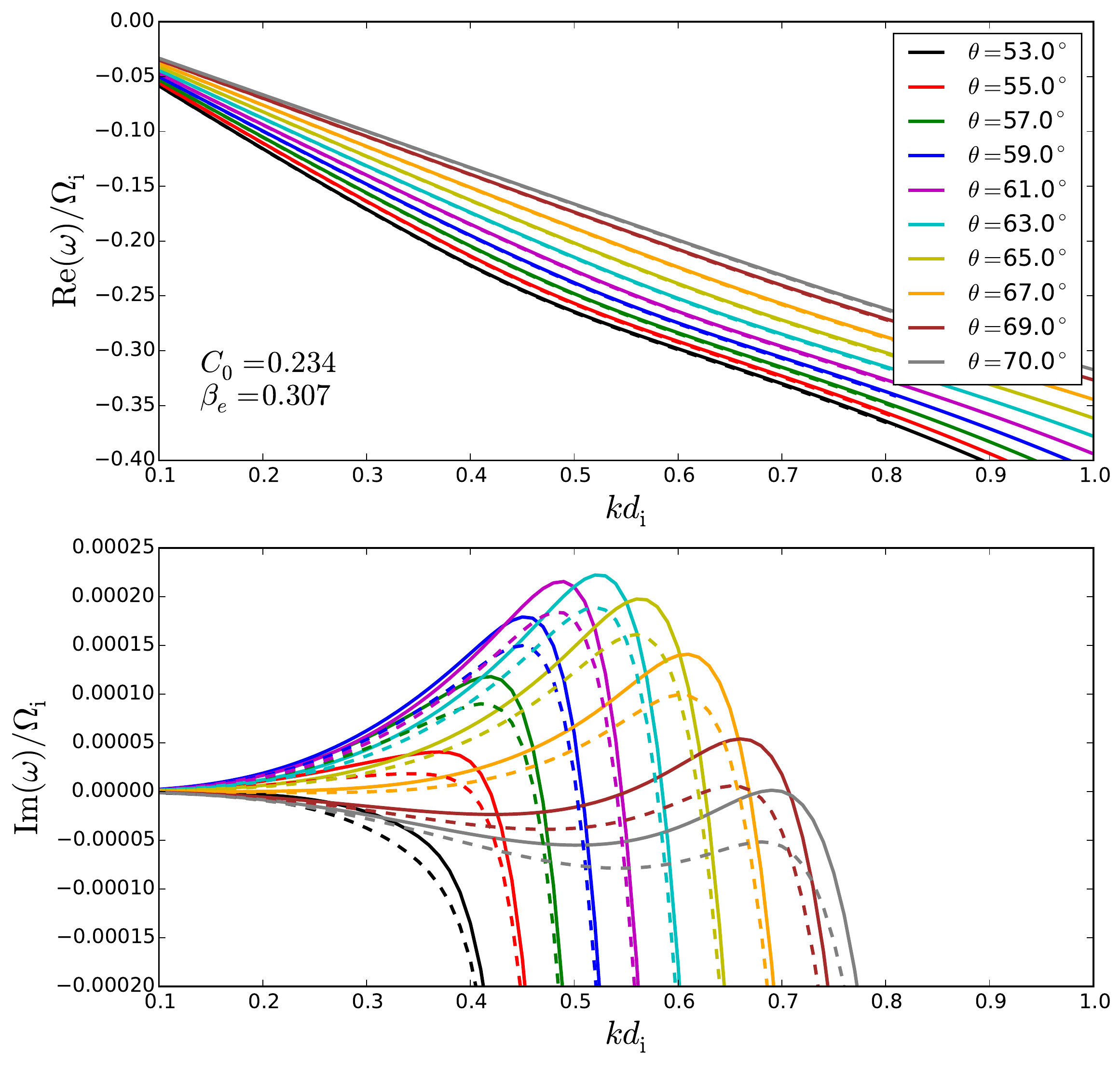}
\caption{\red{ {\it KAW---Less oblique}. Dispersion relations for the original core-strahl model (solid lines, from Fig.~\ref{theta_scan_kaw_plot}) and the core-halo-strahl model (dashed lines) described in this appendix.}}
\label{fig2_comp}
\end{figure}

\begin{figure}
\includegraphics[width=1\linewidth]{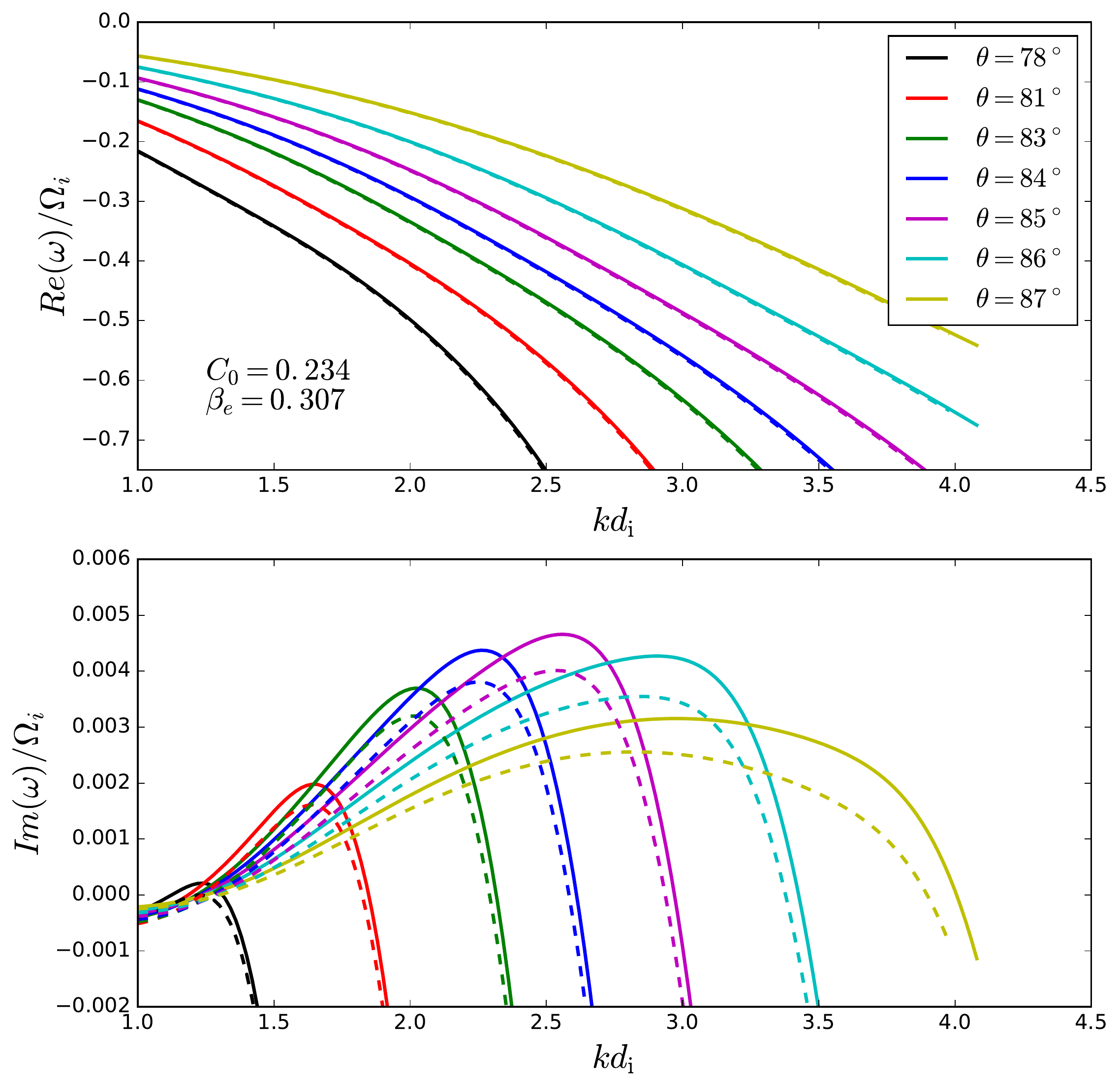}
\caption{\red{ {\it KAW---More oblique}. Dispersion relations for the original core-strahl model (solid lines, from Fig.~\ref{theta_scan_kaw_oblique_plot}) and the core-halo-strahl model (dashed lines) described in this appendix.}}
\label{fig4_comp}
\end{figure}

\begin{figure}
\includegraphics[width=1\linewidth]{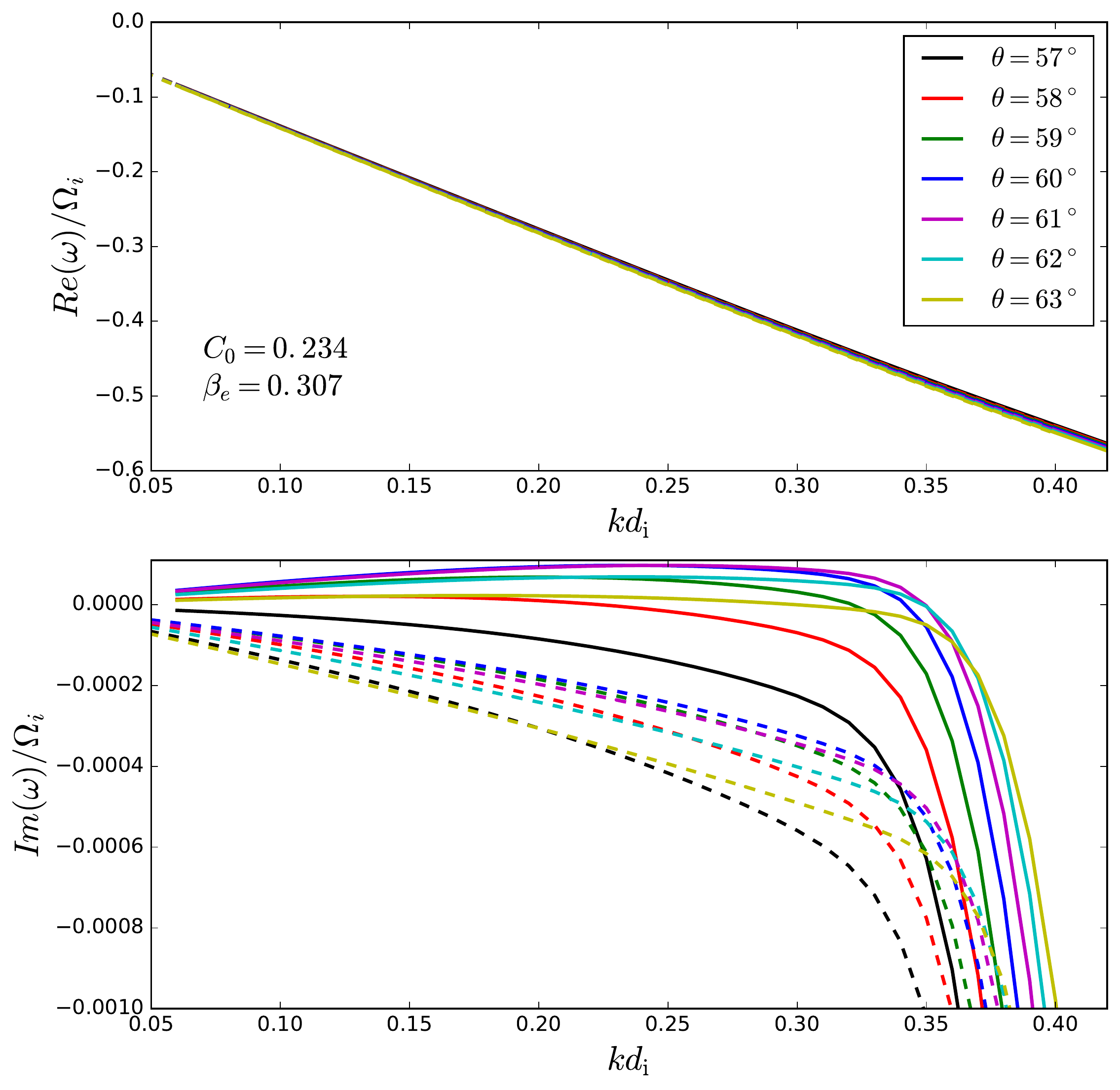}
\caption{ \red{{\it Magnetosonic}. Dispersion relations for the original core-strahl model (solid lines, from Fig.~\ref{theta_scan_ms_plot}) and the core-halo-strahl model (dashed lines) described in this appendix.}}
\label{fig6_comp}
\end{figure}

\begin{figure}
\includegraphics[width=1\linewidth]{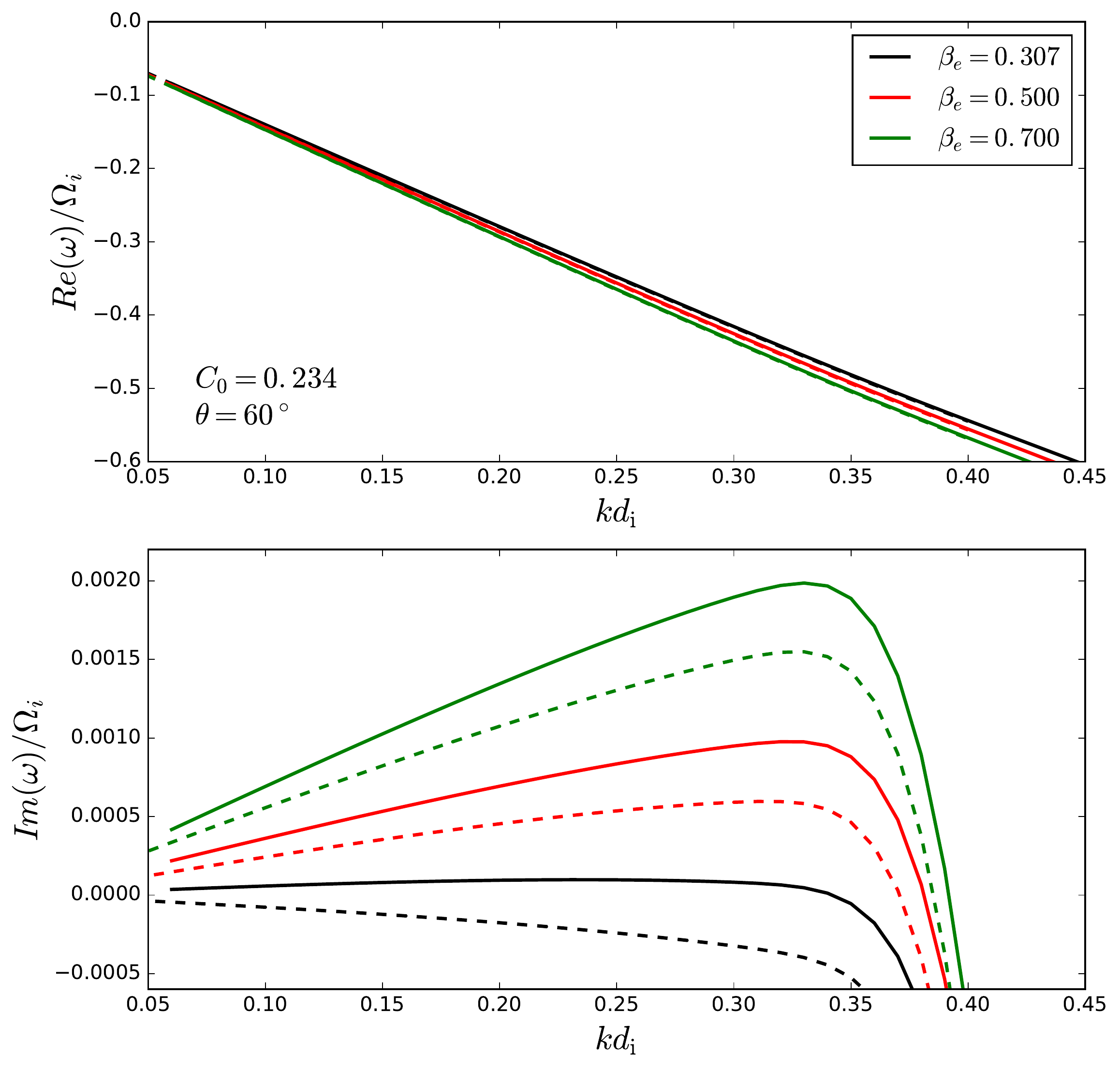}
\caption{\red{ {\it Magnetosonic}. Dispersion relations for the original core-strahl model (solid lines, from Fig.~\ref{beta_scan_ms_plot}) and the core-halo-strahl model (dashed lines) described in this appendix.}}
\label{fig7_comp}
\end{figure}

\red{In Figs.~\ref{fig2_comp}-\ref{fig7_comp}, we compare the dispersion relations for the core-strahl model (solid lines) with those found for the core-halo-strahl model (dashed lines). In all plots, we see that $Re(\omega)$ for the wavemodes is not changed significantly, while the $Im(\omega)$ is made slightly more negative due to the damping introduced by the halo.}

\red{ For the KAW mode, as plotted in Figs.~\ref{fig2_comp}, \ref{fig4_comp}, the character of the instabilities are not significantly altered by the inclusion of the halo. Naturally, the extra damping in the core-halo-strahl case slightly narrows the range of angles over which the instability exists.}

\red{Inclusion of the halo component has a more noticeable effect on the magnetosonic mode in a $\beta_e = 0.307$ plasma, see Fig.~\ref{fig6_comp}. We see that the damping introduced by the halo is significant enough to stabilize this mode at this particular $\beta_e$. However, the magnetosonic instability still exists at higher $\beta_e$ (i.e. at $\beta_e \gtrsim 0.5$), as we can see from Fig.~\ref{fig7_comp}. The halo damping thus has the effect of pushing the instability to a slightly different regime in $\beta_e$.}

\red{As for the core-strahl case, we find that the whistler mode is stable for our core-halo-strahl eVDF. }

\begin{table}
\centering
\begin{tabular}{| c || c |}
\hline
$n_h$  & $n_c/10$\\
$T_h$ & 61.05 eV \\    % = sqrt(\tilde{\gamma} * n / 0.02012)
$\kappa$ & 5 \\
\hline
\end{tabular}
\caption{Here we present the set of physical constants, that appear in Eqs.~\ref{fh_eq} and \ref{ah_eq}, that specify our halo model. }
\label{halo_param_table}
\end{table}

%}

{\em Acknowledgments}---This work of KH and SB was supported by the NSF under the grant no. NSF PHY-1707272 and by NASA under the grant no. NASA 80NSSC18K0646. SB was also supported by the Vilas Associates Award from the University of Wisconsin - Madison.

\bibliographystyle{mnras} % Tell bibtex which bibliography style to use
\bibliography{instability_letter_refs}{}

\end{document}